\documentclass[conference]{IEEEtran}
\usepackage{cite}

\ifCLASSINFOpdf
\else
\fi
\usepackage{array}

\usepackage{stfloats}
\usepackage{url} 

\usepackage[usenames,dvipsnames]{pstricks}
\usepackage{epsfig}
\usepackage{pst-grad} 
\usepackage{pst-plot} 
\usepackage{float}

\usepackage{etoolbox}

\usepackage{hyperref}

\usepackage{multirow}

\begin{document}
%
\title{An end-to-end (deep) neural network applied to raw EEG, fNIRs and body motion data for data fusion and BCI classification task without any pre-/post-processing}

\author{
    \IEEEauthorblockN{
    Aras R. Dargazany\IEEEauthorrefmark{1}\IEEEauthorrefmark{2},
    Mohammadreza Abtahi\IEEEauthorrefmark{1},
    Kunal Mankodiya\IEEEauthorrefmark{1} 
    }
    \IEEEauthorblockA{\IEEEauthorrefmark{1} 
    Department of Electrical, Computer, and Biomedical Engineering,
    University of Rhode Island, USA}
	\IEEEauthorblockA{\IEEEauthorrefmark{2}
    Corresponding Author, arasdar@uri.edu}
}

%


\maketitle

\begin{abstract}
Brain computer interfaces (BCI) using EEG, fNIRS and body motion (MoCap) data are getting more attention due to the fact that fNIRS and MoCap are not prone to movement artifacts similar to other brain imaging techniques such as EEG. 
Advancements in deep learning (neural networks) would allow the use of raw data for efficient feature extraction without any pre-/post-processing. 
In this work, we are performing human activity recognition (BCI classification task) for 5 activity classes using an end-to-end (deep) neural network (NN) (from input all the way to the output) on raw fNIRS, EEG and MoCap data. 
Our core contribution is focused on applying an end-to-end NN model without any pre-/post-processing on the data. 
The entire NN model is being trained using backpropagation algorithm. 
Our end-to-end model is composed of a four-layered MLP: input layer, two hidden layers (using fully connected (dense) layer, batch normalization and leaky-RELU as non-linearity and activation function), and output layer using softmax. 
We have reached minimum 90\% accuracy on the test dataset for the classification task on 10 subjects data and 5 classes of activity. 
\textit{\textbf{The Github link for our detailed implementation and dataset is available here: \url{https://github.com/arasdar/BCI}}}.
\end{abstract}

\textbf{Index Term}: End-to-End neural network, Deep Learning, BCI classification task, Motion Activity Recognition, functional Near-Infrared Spectroscopy (fNIRS), and EEG.

%
\IEEEpeerreviewmaketitle


\begingroup
\section{Introduction}
\label{sec:introduction}

Brain computer interface (BCI) is a direct pathway for communication between brain and external devices. 
It bypasses the neural pathways from brain to different organs. 
In order to develop BCI systems, brain activity recording is needed. 
Functional near-infrared spectroscopy (fNIRS) is one of the non-invasive neuroimaging techniques that utilizes near-infrared (NIR) light in order to measure the blood flow in the cortical regions of brain. 
The layers of the head such as skin, skull, and lipid layers are almost transparent to the NIR light, and the most absorbers of this light are
oxyhemoglobin (HbO2) and deoxyhemoglobin (Hb) in the blood. 
Therefore, fNIRS uses two wavelengths in the NIR range in order to measure the concentration level of HbO2 and Hb by using the Beer-Lambert Law. 
In order for brain to perform a task, it needs oxygen which is carried by blood, and therefore, during the brain activity, there is a demand for more oxygenated blood, which results in increase of HbO2 level \cite{ferrari2012brief,leff2011assessment,villringer1993near,wriessnegger2008spatio,wang2012biomedical}

BCI conventional classification pipeline is composed of a local feature extraction for filtering and denoising purposes (e.g. spatial Laplacian and Morlet Wavelet), global feature extraction, e.g. principle component analysis (PCA) and independent component analysis (ICA), and machine learning approaches, e.g. linear component analysis (LDA) or step-wise linear component analysis (SWLDA) for classification.
An end-to-end neural network (NN) pipeline for the entire BCI classification task can avoid fixed local filtering and denoising, manual local feature extraction, manual data analysis, and different classification approaches. 
This end-to-end approach is also capable of adapting and scaling to bigger datasets as more data become available via more experimental data collection.
If the entire BCI classification pipeline can be replaced with one end-to-end pipeline, then we can conclude that the pre-processing steps for filtering and denoising in BCI classification task is not really necessary.
This explanation of the motivation behind this is visually demonstrated in figure \ref{fig:motive}.
\begin{figure*}[ht]
\centering
\includegraphics[width=\linewidth]{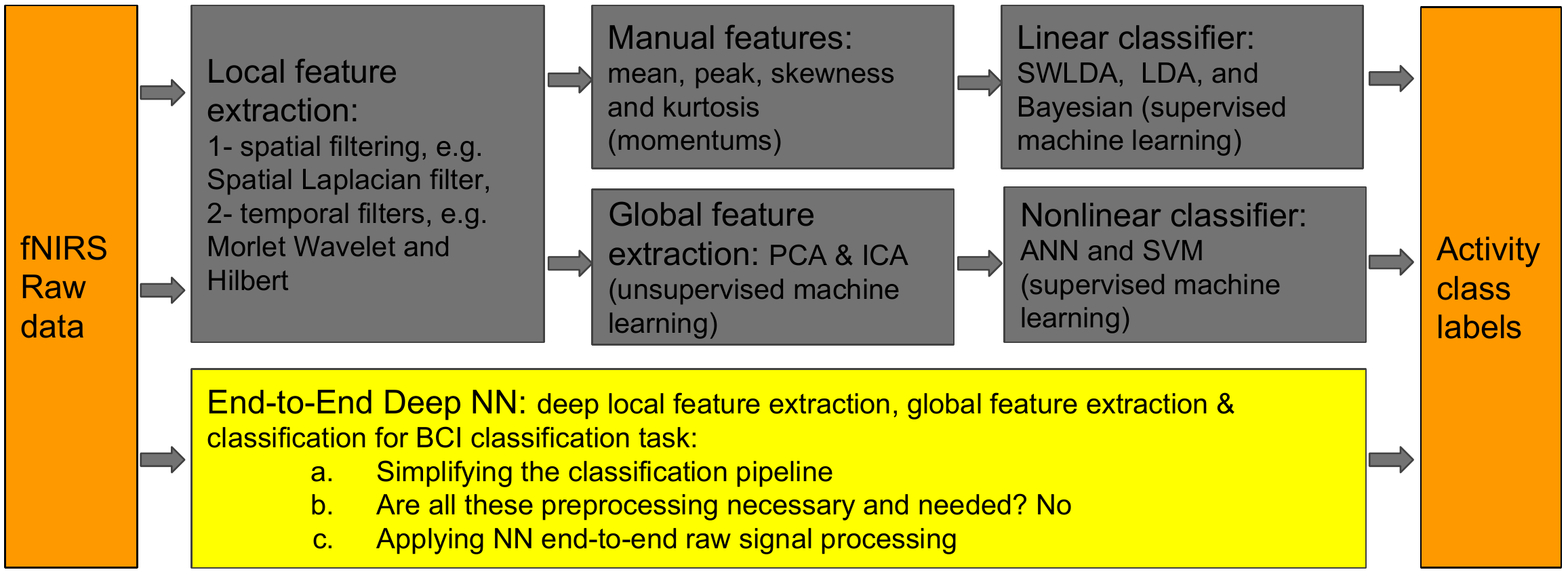}
\caption{Our proposed end-to-end pipeline vs conventional pipeline for BCI classification and recognition tasks.}
\label{fig:motive}
\end{figure*}

In this paper, the aim is to identify different movement activities using the fNIRS data recorded from 10 healthy subjects. 
Conventional classifiers usually need some pre-processing and feature extraction methods to be performed in order to achieve reasonable classification accuracy. 
With advancements in machine learning and use of deep learning algorithms, there is less need for pre-processing or manually extracting features and selecting the best features. 
An end-to-end NN has been used for discriminating different activities.

The remaining of the paper is organized as follow: 
Section \ref{sec:related-works} reviews the related works, 
Section \ref{sec:materials-and-methods} provides details of the data collection and the approach for activity recognition/classification. 
The results and the performance of the identification is demonstrated in Section \ref{sec:results}, 
and Section \ref{sec:discussion-and-conclusion} will conclude the paper.

\section{Related works}
\label{sec:related-works}
Cecotti et al. (in 2008 \cite{cecotti2008convolutional}) applied convolutional neural networks (CNN) to BCI, one of the pioneering work in terms of applying deep learning (DL) to BCI.
He applied CNN with embedded Fourier transform for electroencephalogram (EEG) classification.
In 2011 \cite{cecotti2017convolutional}, he again applied CNN for P300 BCI experimental data for detection purposes.
In another work and roughly the same year \cite{cecotti2011time}, he applied CNN in time-frequency domain for offline classification of steady state event-related potentials (ERP) for recognizing evoked potential responses classes.
Recently (in 2017 \cite{cecotti2017convolutional}), he discussed the power and the impact of CNN architecture in ERP detection.

Trakoolwilaiwan et al. (in 2017 \cite{trakoolwilaiwan2017convolutional}) applied CNN to fNIRS data for move and rest BCI classification task for automating feature extraction and classification modules in conventional pipeline of BCI classification (figure \ref{fig:motive}).
They used a pre-processing step (Wavelet and multi-resolution filters) for denoising the data and then applied a deep CNN for classification.
The reported classification accuracy using CNN is better than support vector machine (SVM) and vanilla artificial neural network (ANN) although the reported online processing time is much worse.
This work \cite{trakoolwilaiwan2017convolutional} inspired/ motivated our work in terms of replacing the pre-processing step with an end-to-end DL pipeline for a fully automated local feature extraction, global feature extraction, and classification in BCI classification task (figure \ref{fig:motive}).
Our motivation and our proposed end-to-end CNN approach (convolutional pipeline) for BCI classification task is illustrated in this figure \ref{fig:motive}.

A study has shown the implementation of a fNIRS-BCI application, \textbf{'Mindswitch'} that harnesses motor imagery for control \cite{coyle2007brain},
and the initial results showed the suitability of fNIRS system to develop simple BCI systems, and has a great potential to be used in more complex systems by developments in classification techniques.
The fNIRS-based  BCI studies between 2004 and 2014 has been reviewed \cite{naseer2015fnirs}.
They have evaluated studies in terms of the experimental tasks, noise removal methods, features, and classification methods. 
It is shown that the most common noise removal method was bandpass filtering, and artificial neural network or ANN, support vector machine or SVM, LDA, and hidden Markov model (HMM) has been used widely for classifying the fNIRS data.
Another study \cite{khoa2008functional} presented the analysis of fNIRS signal and demonstrated the existence of distinct patterns in hemodynamic responses. 
Wavelet analysis and NN has been used for filtering, and classification, respectively.

The use of deep learning on a multi-modal system of combining EEG and fNIRS has been investigated \cite{chiarelli2018deep}, and it is reported that using multi-modal recording and deep neural network (DNN) classifier will result in a significant increase in performance.
Hennrich et al. (2015) \cite{hennrich2015investigating} demonstrated how brain activation patterns measured by fNIRS can be classified by DNN, and compared the results with other common methods. 
Accuracy for classifying different mental tasks has been achieved that are comparable to the results of conventional methods. 

CNN can learn and generalize features automatically, therefore there is no need to define and extract features from the data. 
Hiwa et al. \cite{hiwa2016analyzing} performed CNN analysis on fNIRS data to classify the subject gender. Male and female subject were recruited in this study to perform a visual number memory task in a white noise environment. 
It is suggested that the proposed CNN analysis can be used in order to define regions of interest (ROI) on fNIRS data in order to distinguish features between groups.

Tayeb et al.~\cite{tayeb2019validating} applies deep learning for an end-to-end learning without any feature engineering to BCI motor imagery by combining a spectrogram-based convolutional neural network model (CNN) and long short-term memory (LSTM) as a recurrent neural network (RNN) into convolutional LSTM (a kind of Convolutional RNN) for decoding motor imagery movements directly from raw EEG signals without (any manual) feature engineering. 
\textbf{Tayeb et al.~\cite{tayeb2019validating} demonstrated the successful real-time control of a robotic arm using deep learning-based BCI.}

Reliable signal classification is essential for using
an electroencephalogram (EEG) based Brain-Computer Interface
(BCI) in motor imagery (MI) training. 
While deep learning (DL)
is used in many areas with great success, only a limited number of
works investigate its potential in this domain. 

Dose et al.~\cite{dose2018deep} also applied an end-to-end CNN to raw
EEG signals for spatio-temporal feature extraction and classification. 
The experimental results were superior to the literature on the same data.
With no pre-processing needed, Dose et al.~\cite{dose2018deep} also demonstrate experimentally the power of DL methods as EEG classification methods.

On the other hand, Huve et al. \cite{huve2018fnirs} applied fNIRS-based BCI to assess how driver's mental state is impacted by different external conditions such as weather condition, type of road, including manual driving versus auto-pilot.
An DNN and RNN is applied to fNIRS signals and they both demonstrated the same performance. 

Lacrama et al.~\cite{WWW-sciencemag-ai-turns-brain-activity-speech} is focused on the impact of AI on BCI in the human society in the near future.
The authors believe that special attention should be given to BCI as a possible solution to make man compatible with his future AI environment.
An important example is the ability of current AI tech to convert brain activity signal into speech.
This example and AI-BCI loop/cycle is illustrated in figure~\ref{fig:bci-ai}.
\begin{figure*}
    \centering
    \includegraphics[width=\textwidth]{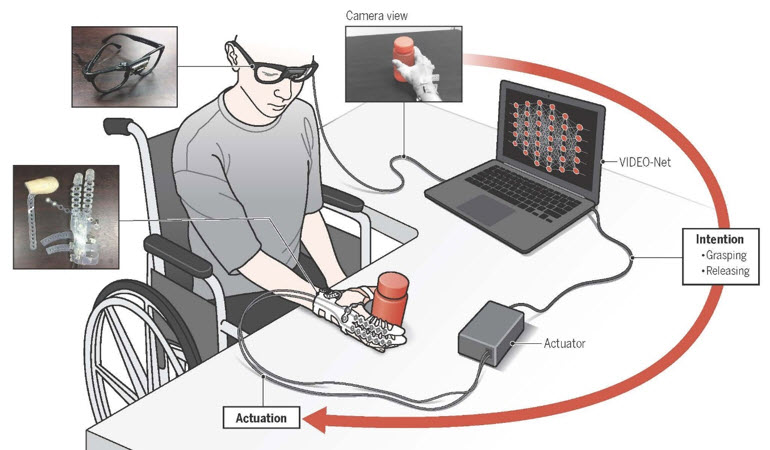}
    \caption{An example of BCI-AI loop: Robot hand uses machine learning to detect wearer’s intention~\cite{WWWW-therobotreport-robot-hand-ml-intention}}
    \label{fig:bci-ai}
\end{figure*}

This work~\cite{angrick2019interpretation} applied an CNN to reconstruct an audible waveform from invasively-measured brain activity. 
They experimentally achieved statistically-significant correlations between spectrogram of synthesized and original speech. 
Their trained models show that electrodes placed in cortical regions associated with speech production tasks have a large impact on the reconstruction of speech segments.

An 1D-Convolutional LSTM~\cite{sun2019eeg} is proposed for EEG-based user identification system whose performance was validated with a public database consists of EEG data of 109 subjects. 
The experimental results showed that this 1D-Convolutional LSTM~\cite{sun2019eeg} has a very high averaged accuracy of 99.58\%, when using only 16 channels of EEG signals, which
outperforms the state-of-the-art EEG-based user identification methods. 
An 1D-Convolutional LSTM~\cite{sun2019eeg} utilizes the spatio-temporal features of the EEG signals with LSTM, and lowering cost of the systems by reducing the number of EEG electrodes used in the systems. 
\section{Materials \& methods}
\label{sec:materials-and-methods}

\subsection{Data Acquisition \& Collection}
In this research, ten healthy participants were recruited, and were consented based on institutional review board (IRB) requirements on IRB No. HU1415-041. 
The NIRScout System (NIRx Inc., New York, NY, USA) with 8 light sources and 8 light detectors has been used. 
The optodes has been set on the motor cortex area on the head, providing 20 channels (attributes) of data. 
Due to the fact that fNIRS uses two different wavelengths in order to detect the changes of HbO2 and Hb, therefore we have 40 channels of data in total for each recording. 
The participants were asked to perform simple movement tasks of moving the arm on right and left hands, moving the leg on right and left legs, and moving both arms together. 
Each activity has been performed for 10 seconds, followed by 20 seconds of rest, and repeated for 5 times, recorded by the sampling rate of 7.8125 Hz.

\subsection{Data Classification}
For data mining and classification of 5 human activities, we use an end-to-end NN on raw fNIRS data. 
This end-to-end NN model without any pre-processing on the raw data is our core focus and contribution. 
The entire NN model is being trained using backpropagation algorithm (Adam). 
This NN model is a four-layered multi-layer perceptron or MLP composed of:
\begin{enumerate}
\item An input layer, 
\item Two hidden layers in which each hidden layer is composed of: 
fully connected (FC) layer, also known as dense layer (as shown in the code snippet figure \ref{fig:model}), 
batch normalization layer,
leaky-RELU which is the non-linear activation function. 
\item Finally the output layer which is composed of a FC layer (dense layer) and a softmax function. 
The softmax function is applied for the loss function in the implementation.
\end{enumerate}

We have reached 77-80\%  accuracy on the classification task on 10 subjects data and 5 classes of activity as shown at the bottom of figure \ref{fig:results}. 
\textbf{\textit{The Github link for the implementation and dataset is available: \textit{\url{https://github.com/arasdar/BCI}}.}}
Our proposed end-to-end NN model for BCI classification task (motion activity recognition) is illustrated in figure \ref{fig:model}.
\begin{figure*}[ht]
\centering
\includegraphics[width=\linewidth]{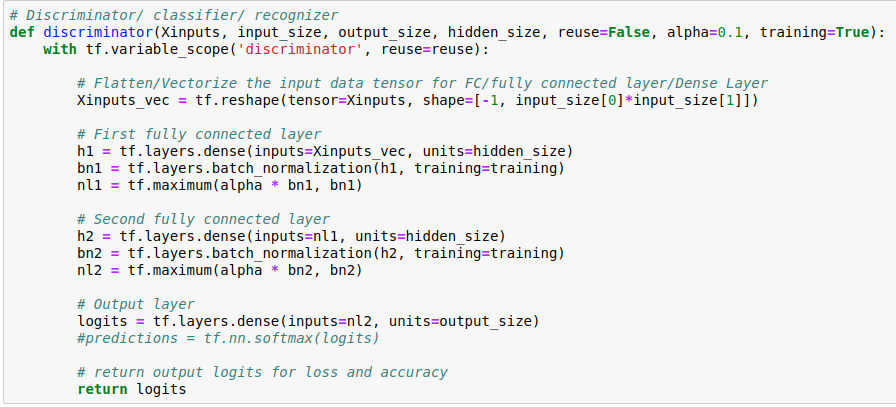}
\caption{This is the code snippet of the model architecture \textit{\textbf{(available on \url{https://github.com/arasdar/BCI} along with the dataset)}} for the implementation details and reproducibility of the results.}
\label{fig:model}
\end{figure*} 
\section{Results}
\label{sec:results}

Our goal is to recognize the activity response in fNIRS data once the stimulus is given the subject. 
Quantifying the stimulus change can be measured in various aspects of the response, such as amplitude, latency, activity location on the scalp and whether this change will affect BCI classification results.
Our Project Plan for data collection was:
1. Recording data from 10 subjects performing 5 activities. Each Subject will perform multiple rest-activity runs.
2. No filtering or pre-processing is done or applied to the collected data before classification. 
No down-sampling or up-sampling is done or applied to the collected data before classification.
The BCI classification task is performed on the raw collected data.

The resulting end-to-end NN pipeline performance for BCI classification task is illustrated in figure \ref{fig:results}.
\begin{figure*}[ht]
\centering
\includegraphics[width=\linewidth]{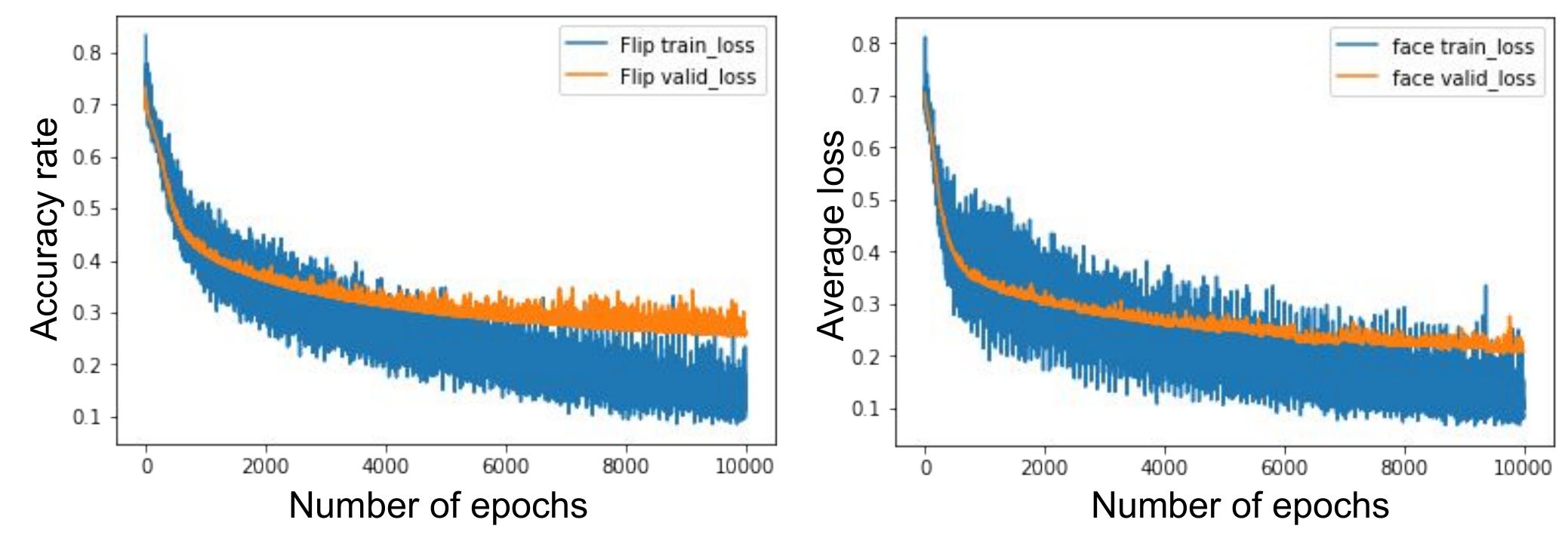}
\includegraphics[width=\linewidth]{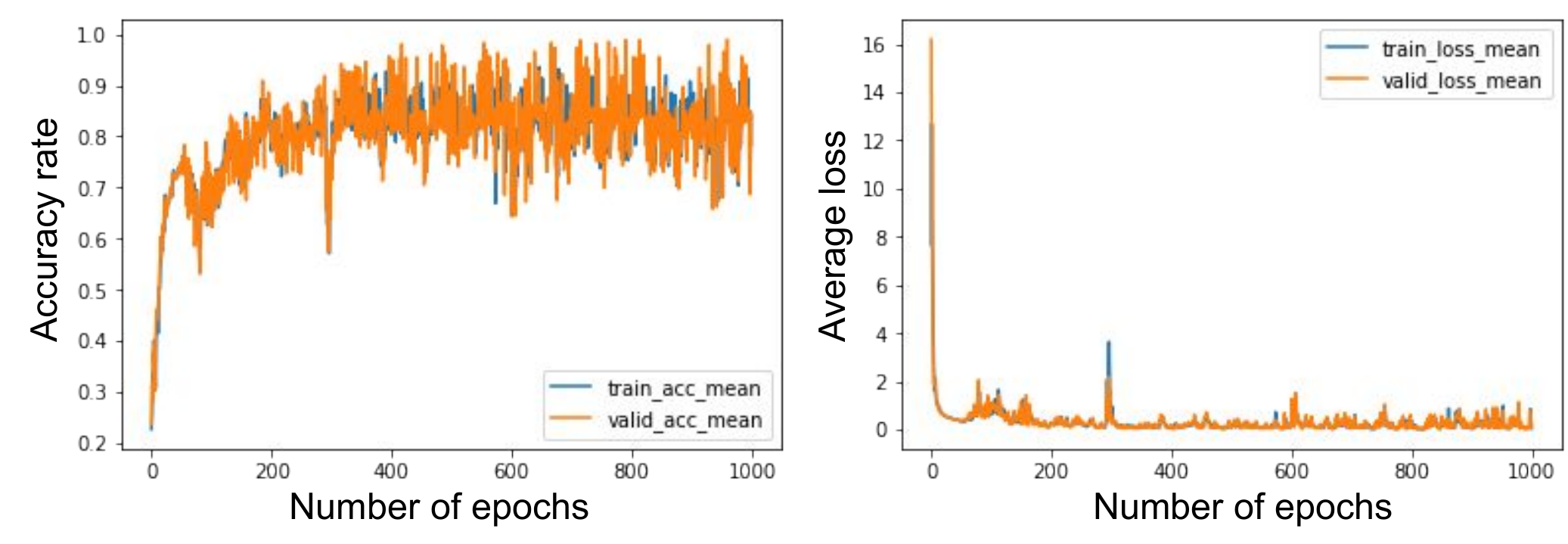}
\includegraphics[width=\linewidth]{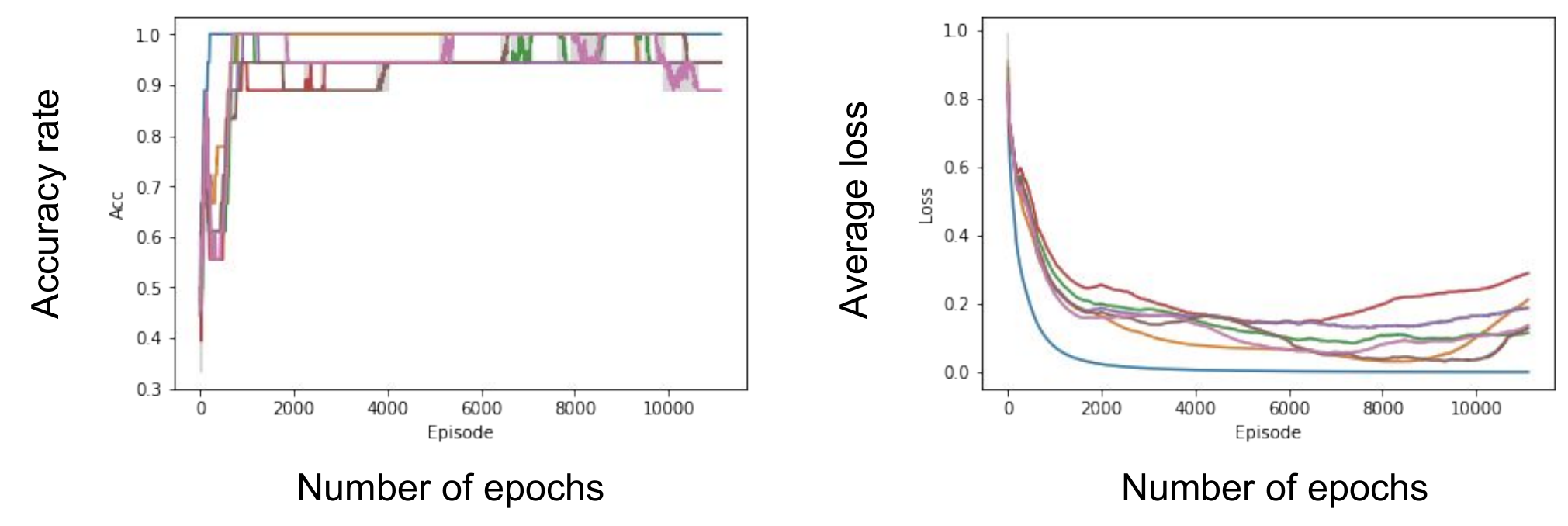}
\caption{Our proposed end-to-end deep NN performance for BCI classification pipeline with minimum 90\% classification accuracy on the test dataset: top row is the result for fNIRS training and validation data; bottom row is the result of applying the model for fusing EEG, MoCap (Motion capture device) and fNIRs on PD and Control}
\label{fig:results}
\end{figure*} 
\section{Discussion \& conclusion}
\label{sec:discussion-and-conclusion}


In this work, we applied an end-to-end NN to a BCI classification task:
\begin{enumerate}
\item To get rid different manually designed modules
\item An automatized learning pipeline
\item More importantly, we showed that pre-processing and denoising are not actually required for BCI classification task
\end{enumerate}
DNN is proved to be a powerful automatic local and global feature extractor with applications to classification. 
We believe the next step should be to investigate how we can adapt and scale this NN to more collected data for making use the data availability from different BCI domain.

We reached the accuracy rate of 77-80\%  using simple MLP on the raw fNIRS data for human activity recognition or HAR (figure \ref{fig:results}).
In future we plan to:
\begin{enumerate}
\item Collect more fNIRS data and increase our dataset size and updated our HAR result using the same model.
\item Compensate the dataset size increase by adding more graphical processing units (GPUs).
\item Apply our model to pre-processed data and compare our results with similar works so that we can validate our resulting accuracy.
\end{enumerate}

\subsection{Pros and cons}
The implemented end-to-end pipeline has the following cons as follows:
\begin{enumerate}
\item NN is computationally heavy (GPU is needed).
\item They are data hungry (need a lot of data).
\item Overfitting is highly possible due to the large number of parameters. 
\end{enumerate}
Having mentioned cons, the proposed end-to-end convolutional architecture has also the following pros:
\begin{enumerate}
\item Makes BCI classification task very simple.
\item Proves CNN useful for signal processing as well as image processing.
\end{enumerate}

\subsection{Future work and perspective}
The NN algorithm/architecture is applicable and highly capable of time-series data processing/learning such as electromyogram (EMG) and other kinds/modality of wearable data such as heart rate.


\endgroup



\end{document}